\documentclass{aa}
\usepackage{graphicx}
\usepackage{txfonts}
\usepackage{times}
\usepackage{epsfig}
\usepackage{graphicx}
\usepackage{amsmath}
\usepackage{amssymb}
\usepackage{tabulary}
\usepackage{multirow}
\begin{document}

\title{Deriving star cluster parameters with convolutional neural networks.}
\subtitle{I. Age, mass, and size}
\author{J. Bialopetravi\v{c}ius\inst{1, 2}, D. Narbutis\inst{1, 2}, V. Vansevi\v{c}ius\inst{1, 2}}
\institute{
Vilnius University Observatory, Saul\.{e}tekio av. 3, LT-10257 Vilnius, Lithuania
\and
Center for Physical Sciences and Technology, Saul\.{e}tekio av. 3, LT-10257 Vilnius, Lithuania \\
\email{jonas.bialopetravicius@ff.vu.lt}}

\abstract
{Convolutional neural networks (CNNs) have been proven to perform fast classification and detection on natural images and have potential to infer astrophysical parameters on the exponentially increasing amount of sky survey imaging data. The inference pipeline can be trained either from real human-annotated data or simulated mock observations. Until now star cluster analysis was based on integral or individual resolved stellar photometry. This limits the amount of information that can be extracted from cluster images.}
{Develop a CNN-based algorithm aimed to simultaneously derive ages, masses, and sizes of star clusters directly from multi-band images. Demonstrate CNN capabilities on low mass semi-resolved star clusters in a low signal-to-noise ratio regime.}
{A CNN was constructed based on the deep residual network (ResNet) architecture and trained on simulated images of star clusters with various ages, masses, and sizes. To provide realistic backgrounds, M31 star fields taken from the PHAT survey were added to the mock cluster images.}
{The proposed CNN was verified on mock images of artificial clusters and has demonstrated high precision and no significant bias for clusters of ages $\lesssim$3Gyr and masses between 250 and 4,000 ${\rm M_\odot}$. The pipeline is end-to-end, starting from input images all the way to the inferred parameters; no hand-coded steps have to be performed: estimates of parameters are provided by the neural network in one inferential step from raw images.}
{}

\titlerunning{Deriving star cluster parameters with CNNs}
\authorrunning{Bialopetravi\v{c}ius et al.}
\maketitle

\section{Introduction}
In recent years, methods using convolutional neural networks (CNNs) have drastically improved various object recognition tasks from natural images, such as object classification and detection. Examples of image recognition competitions where CNNs have recently driven progress include ILVSCR\footnote{ImageNet Large Scale Visual Recognition Competition} \citep{ILSVRC15}, PASCAL VOC\footnote{The PASCAL (Pattern Analysis, Statistical Modelling and Computational Learning) Visual Object Classes} \citep{faster-rcnn} and Microsoft COCO\footnote{Common Objects in Context} \citep{2014arXiv1405.0312L}. As of now, a lot of these tasks can be solved with a better accuracy than that of a human \citep{ILSVRC15}. These methods are trained wholly on data and forgo any manual feature engineering steps that were usually required for computer vision applications.

The uptake of deep learning based methods has also been accelerating in the field of astronomy. Most of the activity is in galaxy classification \citep{dieleman, 2018ApJ...858..114H, 2018MNRAS.476.3661D}, gravitational lensing \citep{2017arXiv170207675P,2018MNRAS.473.3895L,2018ApJ...856...68P} and transient detection \citep{deephits, 2018MNRAS.473.3895L, 2017arXiv171001422S}. There has also been work on astronomical image reconstruction \citep{2016arXiv161204526F}, exoplanet identification \citep{2018AJ....155...94S}, point spread function modeling \citep{2018arXiv180107615H} and other topics. However, star cluster parameter estimation has not yet been attempted with these methods, even though the field is an ideal candidate since accurate inference from imaging data is sorely needed.

One of the benefits of CNN architectures is that parameter inference tasks can be solved in much the same way as object type classification. This opens up the possibility to perform astrophysical parameter inference from star cluster images both accurately and efficiently, utilizing all of the available information in each pixel of an image. However, the method first has to be trained on either human-annotated or simulated mock observations.

The Panchromatic Hubble Andromeda Treasury (PHAT) \citep{2012ApJS..200...18D} provides high quality multi-band imaging data\footnote{https://archive.stsci.edu/prepds/phat/} along with human-annotated star cluster catalogs \citep{2012ApJ...752...95J, 2015ApJ...802..127J}. These catalogs have previously been analysed by \cite{2014ApJ...786..117F} and \cite{2017A&A...602A.112D}, who provided cluster age and mass estimates using integrated photometry. Based on resolved star photometry, \cite{2017ApJ...839...78J} derived cluster ages and masses. Previously, \cite{2009AJ....137...94C} had derived cluster parameters of massive clusters in M31 based on spectroscopy. Sizes were estimated from images using aperture photometry by \cite{2015ApJ...802..127J}. Therefore, the PHAT dataset provides a good basis for testing CNN-based methods.

In this paper, we propose a CNN architecture to estimate the ages, masses, and sizes of star clusters. The network is trained on realistic mock observations, with backgrounds taken from the PHAT survey. The method is tested on a different set of artificial clusters. The paper is organized into the following sections: \S \ref{sec:data} gives details about the PHAT survey data, \S \ref{sec:deep_learning} gives a summary of used deep learning based methods, \S \ref{sec:method} describes the proposed CNN and its training methodology, \S \ref{sec:results} presents the results of testing the method on artificial clusters.

\section{Data} \label{sec:data}
The PHAT survey data is extensively described by \cite{2012ApJS..200...18D}. We have used stacked, defect-free mosaic ("brick") images, which are photometrically (pixel values are in counts per second) and astrometrically (with available world coordinate system information) calibrated. Multi-band images for four bricks (19, 20, 21, 22) were obtained from the PHAT archive\footnote{https://archive.stsci.edu/prepds/phat/}. The bricks were chosen to be located in the outer part of M31 disk for low stellar background contamination. Although the PHAT survey was conducted in six passbands, only three (F336W, F475W, F814W) were used in this study because of their high signal-to-noise ratio. The frames in these passbands were observed with WFC3 and ACS instruments, and have different PSFs as well as exposure times.

Each passband image of a brick was transformed into a tangential projection with a common scale (0.05 arcsec/pixel) and size, oriented in such a way that the pixel grids of the images are aligned with the north and east directions. First, the world coordinate systems of images were aligned and then pixel values were transformed using the reproject\footnote{https://reproject.readthedocs.io/} package from Astropy\footnote{http://www.astropy.org/}, conserving flux. Bicubic interpolation was used, providing appropriately resampled images for CNN training purposes.

The brick images contain a lot of saturated stars and extended objects, which could introduce bias to the CNN training procedure. To deal with this, we masked out real clusters and galaxies listed in catalogs by \cite{2012ApJ...752...95J, 2015ApJ...802..127J} as well as stars brighter than 18 mag in the G passband of the Gaia catalog \citep{2016A&A...595A...2G}. These regions were omitted when generating backgrounds for artificial clusters.

To generate mock clusters of various ages, masses, and sizes, PARSEC isochrones\footnote{http://stev.oapd.inaf.it/cgi-bin/cmd}, release 1.2S \citep{2012MNRAS.427..127B}, were used to sample stellar masses according to the \cite{2001MNRAS.322..231K} initial mass function. A fixed metallicity of $Z=0.009$ and no interstellar extinction was assumed. From the isochrones the absolute magnitudes were taken, stars were placed at a distance of M31 \citep[785 kpc]{2005MNRAS.356..979M} and their fluxes were transformed to HST counts per second using HST calibrations for ACS \citep{2017acsi.book.....A} and WFC3 \citep{2012wfci.book.....D} cameras.

The spatial distribution of stars was sampled from the Elson-Fall-Freeman (EFF) model \citep{1987ApJ...323...54E}, placing the cluster in the center of the image.

\begin{figure}
    \centering
    \includegraphics[width=1.0\columnwidth]{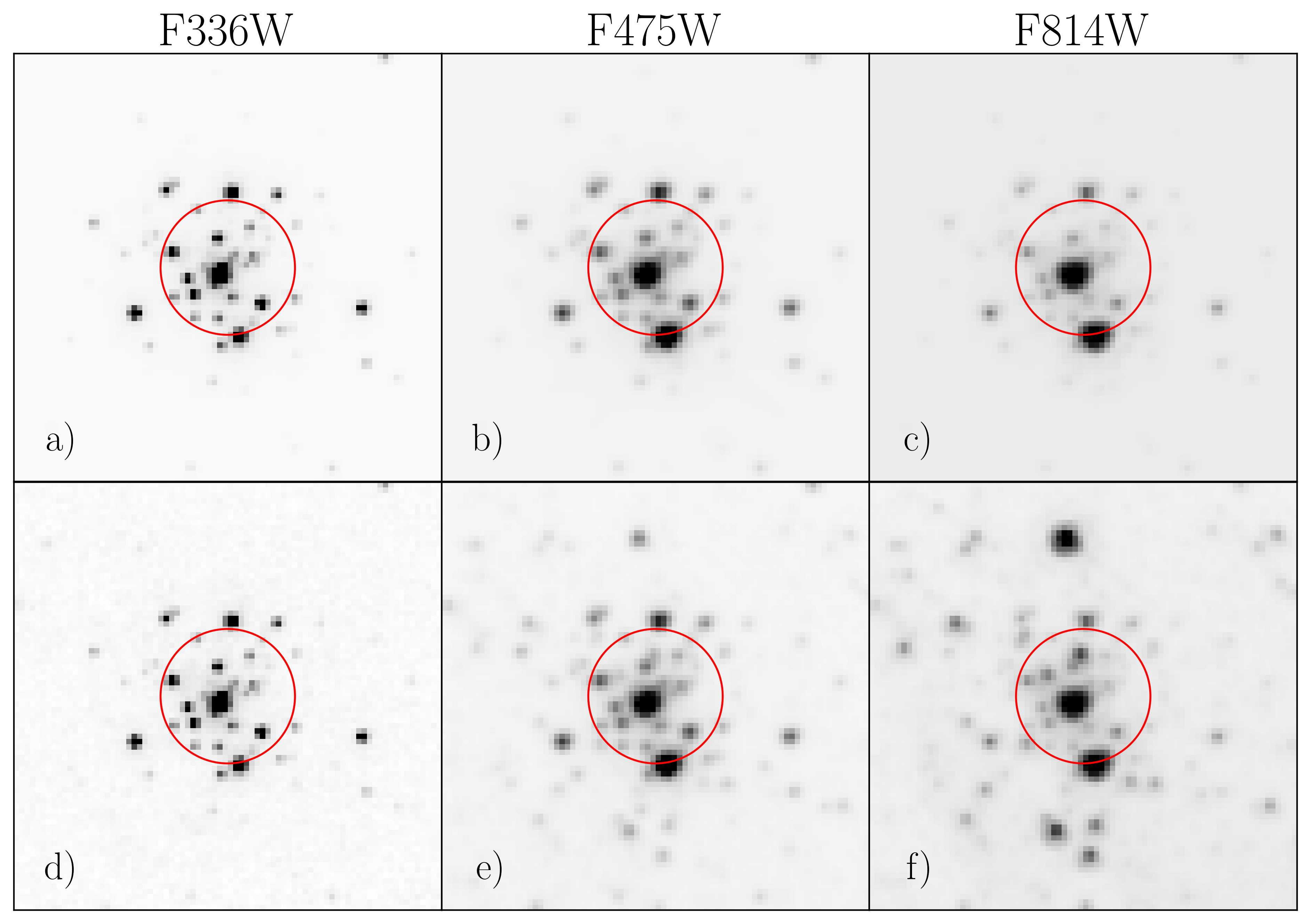}
    \caption{An example of a mock cluster drawn with GalSim, without (top) and with a random background from PHAT (bottom). Passbands are indicated on top of each panel. A cluster with typical parameters of $\log_{10}(t/{\rm yr})=8.0$, $\log_{10}(M/{\rm M_\odot})=3.0$, and $r_h=0.6$ arcsec is shown. The red circle of $r_h$ radius encloses half of the cluster's stars. The intensity of the images is normalized with the arcsinh function.}
    \label{fig:generated_example}
\end{figure}

TinyTim\footnote{http://tinytim.stsci.edu/cgi-bin/tinytimweb.cgi} PSFs \citep{2011SPIE.8127E..0JK} were used to draw the individual stars of the clusters. The PSFs used were 6$\times$6 arcsec in size and drawn with the GalSim package \citep{2015A&C....10..121R} in the coordinate system of aligned PHAT bricks. Artificial clusters were then placed on backgrounds taken from the transformed PHAT bricks. An example of a mock cluster without and with a background is shown in Fig. \ref{fig:generated_example}.

\section{Deep learning} \label{sec:deep_learning}
\subsection{Artificial neural networks}
For a comprehensive review of artificial neural networks (ANNs) and the algorithms for their training, we refer the reader to \cite{haykin2009neural}. Here, a short summary is presented as a basis for the methods used in this study.

The convolutional neural network used in this paper is a type of ANN, which is a machine learning model comprising multiple artificial neurons, each taking a vector of inputs and producing a single output. The vector of inputs is multiplied by a vector of weights, summed and then passed through a non-linearity, called the activation function. One of the most common activation functions is the rectified linear unit (ReLU), defined as $f(x)=max(0,x)$ \citep{Nair:2010:RLU:3104322.3104425}.

An ANN can be trained using the gradient descent algorithm. This requires training samples, which consist of inputs to the network (or observations), and the expected outputs (targets). A loss function $\mathcal{L}$ is used to determine how well the outputs of the network match the expected outputs in the training set. The gradient of this loss function is computed for each sample with respect to the weights of the neural network, and then the weights are adjusted according to the gradient. In essence, the network's weights $w_{t+1}$ at iteration $t+1$ are updated from the weights $w_t$ at iteration $t$ by the formula $w_{t+1}=w_t-\eta\nabla\mathcal{L}$, where $\eta$ is called the learning rate and is used to control the speed of the learning process. Large values of the learning rate result in quick learning, but the model converges poorly due to large parameter steps near optima; meanwhile, small learning rates have the opposite effect. It is important to control this variable during training so that the training procedure converges.

In the most common form of an ANN, neurons are arranged in multiple layers, where each layer has some number of neurons taking inputs from the previous layer and producing activations for the next layer to process. Such an arrangement is called a feed-forward network. The algorithm used for optimizing such networks is called backpropagation, which provides a convenient way to calculate gradients at each layer using the chain rule. With this algorithm, the inputs are passed through the network, obtaining outputs. Gradients are then calculated and propagated backwards, adjusting the weights layer by layer so that the training error (loss) is minimized.

Since each neuron performs a non-linear mapping of its inputs, stacking many layers of neurons results in a deeply non-linear model, which is beneficial in many real world tasks. However, backpropagation-based algorithms often result in gradients vanishing and learning stalling in the lower layers, or in gradients exploding, which results in divergence. Usually, these problems are solved by either minimizing the number of optimizable parameters with more restrictive neural network architectures, or restricting the possible values of the network's weights that can be obtained during training.

\subsection{Convolutional neural networks}
A convolutional neural network is one way to regularize the weights of a neural network via architectural constraints. In a regular ANN, each neuron takes input from every neuron in the lower layers. Meanwhile, in a CNN each layer consists of learnable convolutional filters with a small number of parameters that exploit the 2D structure of images. Each convolutional filter may be as small as a 3$\times$3 matrix, resulting in 9 optimizable parameters. The filter is applied many times to its input by moving it by a step, called a stride, and a lot of outputs for the next layer are produced. Training such models results in increasingly more abstract feature detectors in the form of convolutional filters. The lower layers look for simple features, such as corners or edges of objects. Deeper layers combine these features to form more abstract concepts about objects present in an image, while discarding irrelevant information and noise. It seems that this hierarchical pipeline of feature extraction is essential to solving computer vision problems, as the neocortex of animals is known to work in a similar way \citep{primate_visual_cortex}.

In classical computer vision, hierarchy was usually implemented by hand. Simple features were first detected and extracted, then combined into more complex aggregates, that can describe whole objects, and later processed with simple machine learning algorithms. These hand-engineered approaches, however, had subpar performance as real world data has a lot of noise, irregularities, irrelevant correlations and a high level of variation. Meanwhile, CNNs learn the necessary regularities from the data itself without any feature engineering. This is also why CNNs are such a favorable algorithm to apply to star clusters.

\subsection{Residual networks}
During the past few years, many variants of CNN architecture have been proposed. It has become quite common in literature to reuse these standardized architectures in order to have a common ground of reference for measuring improvements.

\begin{figure}
    \centering
        \includegraphics[width=0.95\columnwidth]{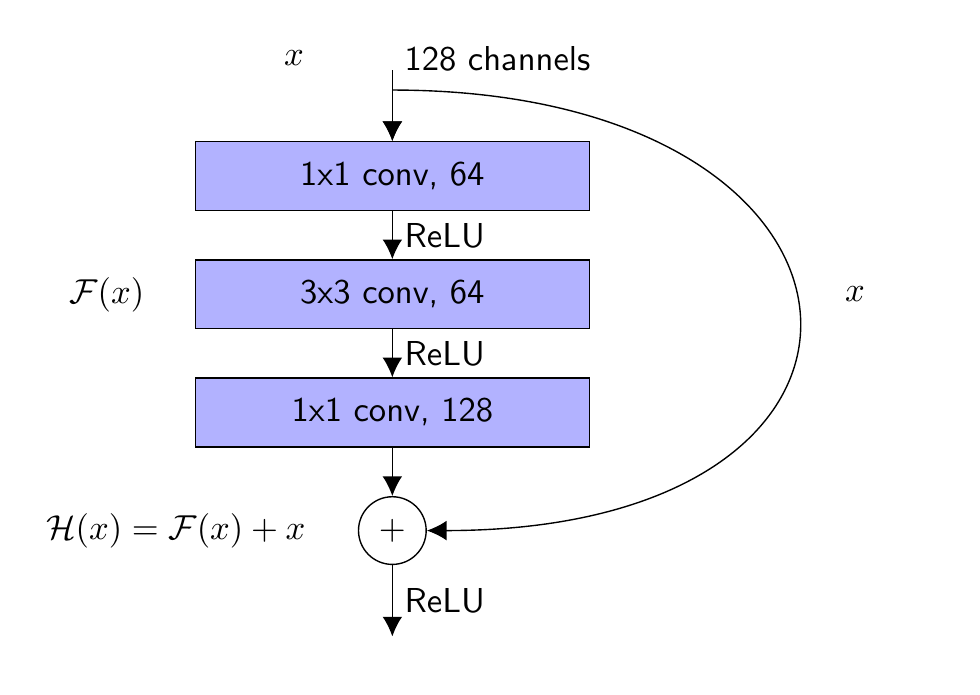}
    \caption{Example building block of a residual network (ResNet), consisting of three sequential convolutional layers. The input is a 128-channel activation map, which is passed through 64 1$\times$1 convolutional filters. The filters extract a 64-channel feature map. These features, after applying a ReLU activation, are then passed through 64 3$\times$3 convolutional filters. The purpose of the first layer is to compress the channels for the 3$\times$3 convolutional layer, which results in less optimizable parameters. Then, the ReLU activation is applied again and the final 1$\times$1 convolutional layer expands the number of channels back to 128. Finally, these outputs are summed with the inputs via a skip-connection and passed through a ReLU activation.}
    \label{fig:resnet_block}
\end{figure}

One such architecture is called the ResNet \citep{2015arXiv151203385H}. Convolutional layers in a regular CNN learn feature detectors, taking inputs from lower layers and mapping them to a representation, called a feature map. The idea of a ResNet is that instead of learning to produce a simple feature map, a residual mapping on top of the features of the previous layer is learned. Denoting the inputs of a layer as $x$, we're looking for a mapping $\mathcal{H}(x)$, such that $\mathcal{F}(x)=\mathcal{H}(x)-x$ (the residual function). This gives the desired mapping as $\mathcal{H}(x)=\mathcal{F}(x)+x$. This has a convenient implementation for CNNs as a skip-connection, which consists of simply taking the activations of a layer and adding it to a deeper layer in the network's graph (see Fig. \ref{fig:resnet_block}).

The ResNet architecture has allowed \cite{2015arXiv151203385H} to reach state-of-the-art results on a few standard image recognition datasets with a lower number of network weights than the alternatives at the time. The training of ResNet type networks is stable regardless of network depth since good results are achieved with networks as shallow as 20 layers (ResNet-20) and as deep as 1202 layers (ResNet-1202). Such networks also seem to work well on object classification as well as detection and regression tasks. Therefore, they are ideal candidates for application in clusters.

\section{Methods} \label{sec:method}
\subsection{CNN architecture}
\begin{figure}
    \centering
        \includegraphics[width=0.7\columnwidth]{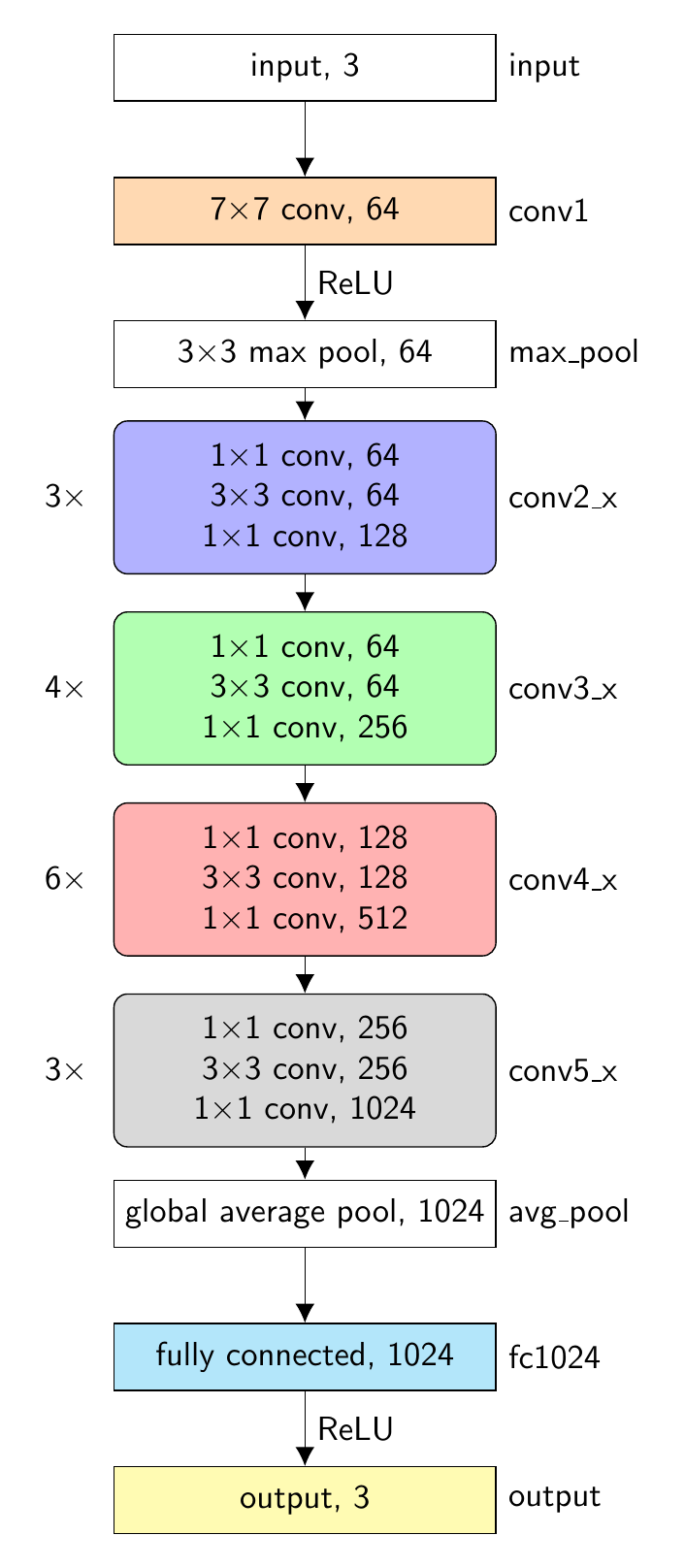}
    \caption{A block diagram of the used CNN. The input image of a cluster passes through the network top to bottom, with the output result being age, mass, and size. All blocks with sharp corners depict singular layers, while blocks with rounded corners are groupings of layers (see example in Fig. \ref{fig:resnet_block}), with the number on the left indicating how many times the group is repeated sequentially and the name on the right corresponding to the layer names in Fig. \ref{fig:network_architecture_table}. The blocks in color are parts of the network with optimizable parameters. The last number in each row is the number of output channels from that layer.}
    \label{fig:network_architecture_graph}
\end{figure}

\begin{figure}
    \centering
        \includegraphics[width=0.8\columnwidth,trim={6.8cm 18.5cm 6.8cm 1.5cm},clip]{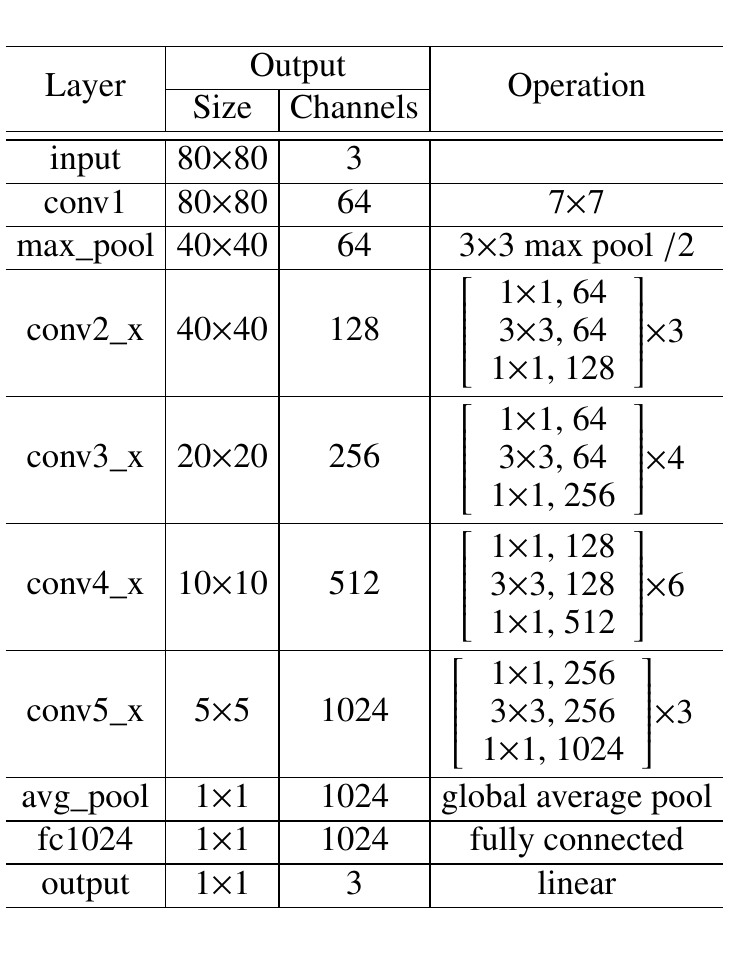}
    \caption{The designed 50-layer CNN, based on the ResNet architecture. The network's layers are listed top to bottom, starting from the images of clusters and with the final layer producing the cluster's age, mass, and size. The convolutional layers are actually groups of blocks depicted in Fig. \ref{fig:resnet_block}, with the "\_x" in the name standing as a placeholder for the block number. The size of the outputs of each layer, both in spatial dimensions and in channel count, are listed on the second and third columns. The last column lists the operations that each layer performs. The layers or blocks with a stride of 2 are: max\_pool, conv3\_1, conv4\_1, and conv5\_1; as can be seen when input and output sizes differ by 2.}
    \label{fig:network_architecture_table}
\end{figure}

In this paper, the ResNet-50 version was used as a basis for the constructed CNN. The network was adapted from the variant used for the ImageNet dataset \citep{ILSVRC15}. The usual inputs of this network are 224$\times$224 pixel natural photographic RGB images (3 channels), which get compressed very quickly in the lower layers to narrow feature mappings, because of the large stride by which the convolutional kernels are moved. Meanwhile, our inputs are 80$\times$80 pixels in size (3 channels: F336W, F475W, F814W), but we operate in a low signal-to-noise ratio regime. Therefore, we reduced the stride of the earliest convolution operations, which should allow the network to extract more low-level information. The network was implemented with the TensorFlow package\footnote{https://www.tensorflow.org/} and is depicted in Figs. \ref{fig:network_architecture_graph} and \ref{fig:network_architecture_table}.

Considering that star clusters have rotational symmetry, rotating an image should not impact what the network infers about it. One way to alleviate this problem is with a method proposed by \citet{dieleman}: the original image of a cluster is rotated by 90 degrees 3 times and passed through the same convolutional layers. The resulting layer activations are then averaged before the output of the network (2nd block from the bottom in Fig. \ref{fig:network_architecture_graph}). This enforces the idea that a rotated image of a cluster should result in the same high-level representations in the network's activations, as the parameters do not depend on the rotation angle. The outputs of the network were represented as a separate neuron for each parameter (age, mass, and size).

\subsection{CNN training} \label{training}
The training procedure consists of propagating the images through multiple layers of convolutions until the outputs (in our case -- the three cluster parameters) are obtained. These outputs are directly determined by the network's structure and weights. The structure is fixed beforehand; however the weights start out random and need to be optimized in order to construct an accurate inferential model, mapping images of clusters to their parameters.

At each iteration of the training procedure, a batch of 64 clusters (which includes all 3 passband images and their 4 rotated variants) is taken and propagated through the network. The inferred output parameters are then compared to the expected outputs and all the weights are updated according to a loss function:
\begin{equation}
\mathcal{L}(targets, outputs)=\sum_{i=1}^3 {\rm smooth_{\mathcal{L}_1}}(targets_i - outputs_i),
\end{equation}
where $targets$ refers to the true values of a given cluster's parameters, while $outputs$ refers to the parameters given by the network; $i$ indicates parameter number (age, mass, and size).

For comparing these values and deriving the training gradient for each of the parameters, the following function was used:
\begin{equation}
{\rm smooth_{\mathcal{L}_1}}(x) = \left\{ \begin{array}{rl}
 0.5x^2 &\mbox{ if $|x|<1$,} \\
  |x|-0.5 &\mbox{ otherwise,}
       \end{array} \right.
\end{equation}
which is a robust Manhattan distance based loss function \citep{2015arXiv150408083G}; it is more resilient to large prediction outliers than the more commonly used mean-squared-error loss. The Adam optimizer \citep{2014arXiv1412.6980K} was used to compute the network's weight gradient updates.

\begin{figure}
    \centering
        \includegraphics[width=0.9\columnwidth]{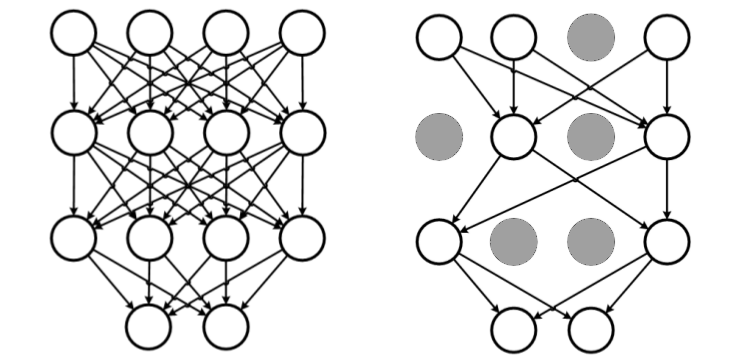}
    \caption{Illustration of the dropout method \citep{dropout}. The left panel shows a fully connected feed-forward neural network, while the right panel shows the same network with some of the neurons disconnected by dropout.}
    \label{fig:dropout_illustration}
\end{figure}

The constructed CNN has $\sim$7 million parameters. Even with a large number of training samples, optimizing this many parameters is problematic. Memorizing the images or unimportant peculiarities in the training data can become much easier than learning actual generalized tendencies that produce the output parameters we seek, resulting in overfitting. To combat this, we used the dropout method \citep{dropout}, which discards neurons randomly during training from the network with probability $p_{\rm dropout}$. An example of this can be seen in Fig. \ref{fig:dropout_illustration}. This reduces overfitting by preventing neurons from adapting to each other, in effect making the network more robust to missing outputs from neurons in other layers. This was only done during training with $p_{\rm dropout}=0.5$, in the last fully connected layer seen in Fig. \ref{fig:network_architecture_graph} (2nd block from the bottom). For inference, the fully trained network with all of the neurons is used.

\subsection{Artificial clusters} \label{artificial_clusters}
\begin{figure}
    \centering
    \includegraphics[width=0.85\columnwidth]{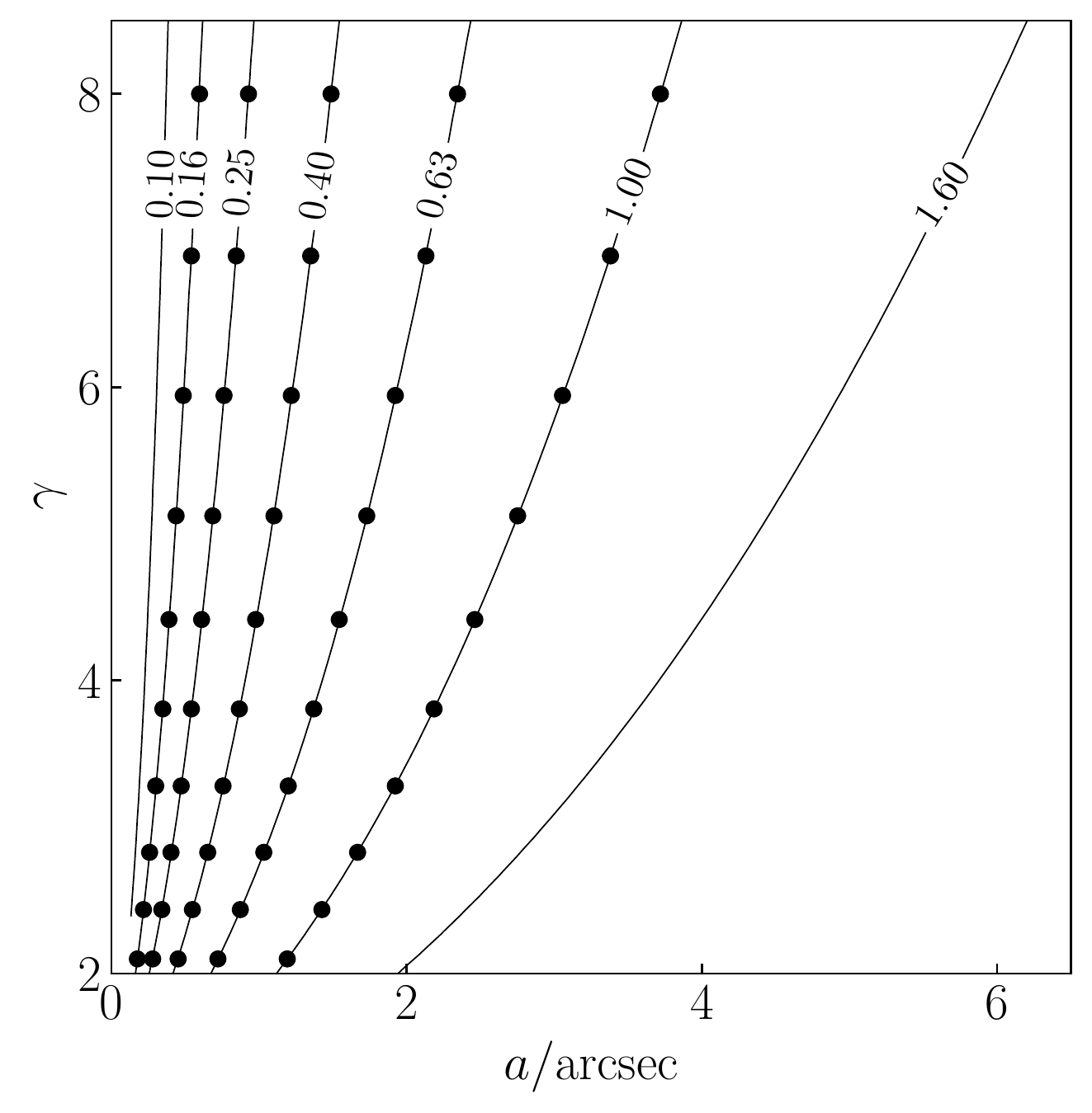}
    \caption{Used EFF parameter space and corresponding lines of constant $r_h$. Clusters for CNN training were sampled from lines, while dots correspond to the values of $r_h$ used to construct a grid of the artificial test samples.}
    \label{fig:eff_rad_construction}
\end{figure}

Each cluster's age was sampled from the logarithmic range of $\log(t/{\rm yr})=[6.6, 9.5]$ (with a step of 0.05 dex); mass was sampled from the logarithmic range of $\log(M/{\rm M_\odot})=[2.0, 4.0]$ to cover the majority of real young, low-mass M31 clusters studied by \cite{2017A&A...602A.112D}. The cluster's star count surface density radial profile $\mu(r)$ was sampled from the EFF \citep{1987ApJ...323...54E} model:
\begin{equation}
\mu(r)=\mu_0(1+r^2/a^2)^{-\gamma/2}.
\end{equation}
The parameters $a=[0.05, 6.4]$ arcsec and $\gamma=[2.05, 8.0]$ were sampled in a logarithmic space within the curves defined by constant $r_h$ values (between 0.1 and 1.6 arcsec), as shown in Fig. \ref{fig:eff_rad_construction}. We define $r_h$ as the radius of a circle on the sky enclosing half of a cluster's stars. These values at the assumed M31 distance (785 kpc) roughly correspond to real cluster sizes in M31 \citep{2009ApJ...703.1872V,2014A&A...569A..30N}. The setup was chosen to target low mass semi-resolved star clusters in order to demonstrate the CNN's capabilities in low signal-to-noise ratio conditions.

Samples of artificial clusters were generated with these parameters and combined with backgrounds of M31 stars. The process is as follows. A random image of a background is selected and its median value is determined. This median is then added to the image of an artificial cluster and then each pixel is sampled from a Poisson distribution, with its mean set to the pixel's value. The median is then subtracted back from this image and the real background image is added. The counts per second of the images were then transformed with the logarithmic function $\log(x + 1)$. The resulting images were 80$\times$80 pixels in size, which correspond to 4$\times$4 arcsec, or 15$\times$15 pc at the distance of M31 (785 kpc).

\begin{figure*}
    \centering
    \begin{tabular}{@{}c@{}}
        \includegraphics[width=0.48\textwidth]{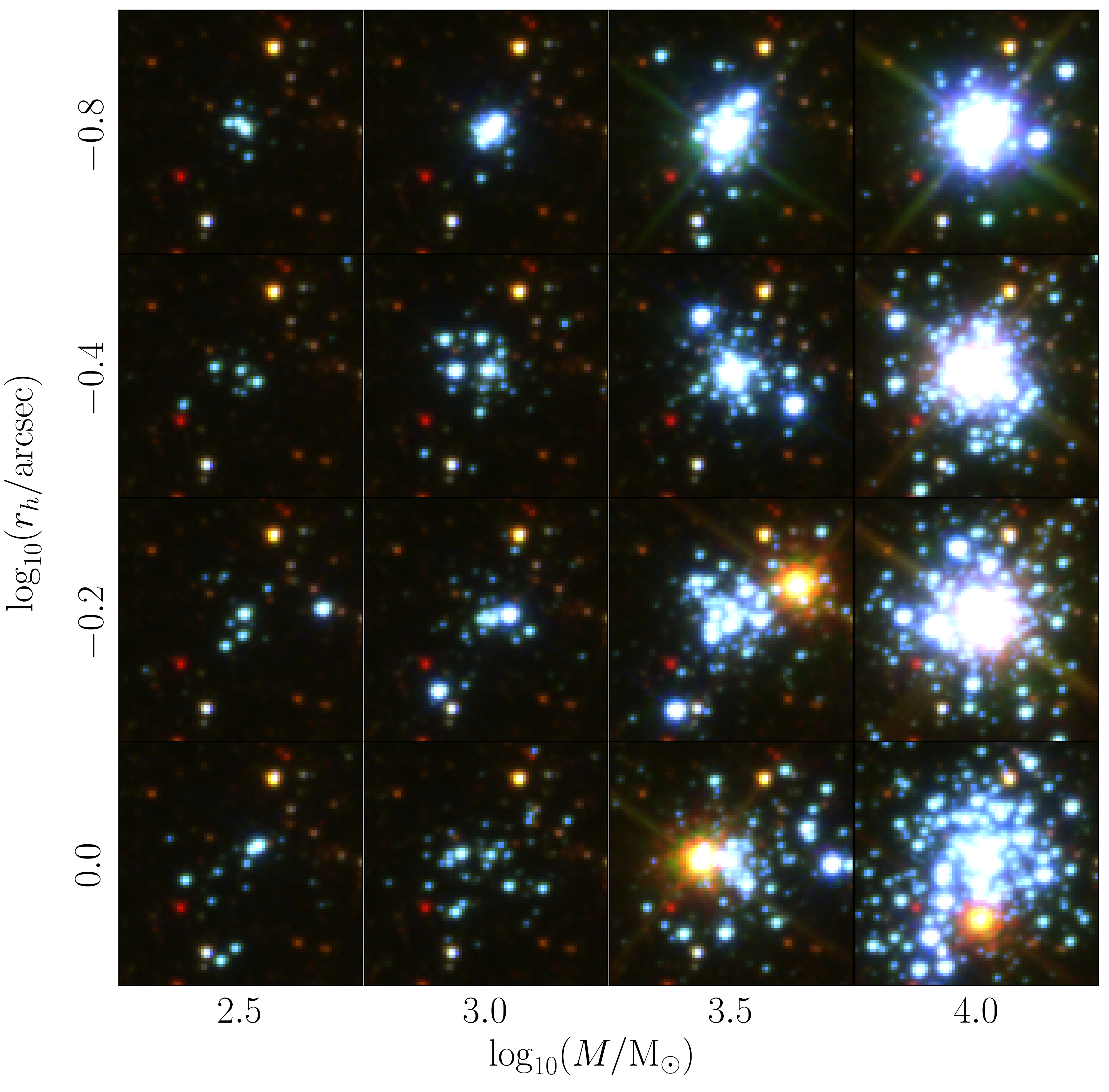} \\[0.5 pt]
        \small a)
    \end{tabular}
    \begin{tabular}{@{}c@{}}
        \includegraphics[width=0.48\textwidth]{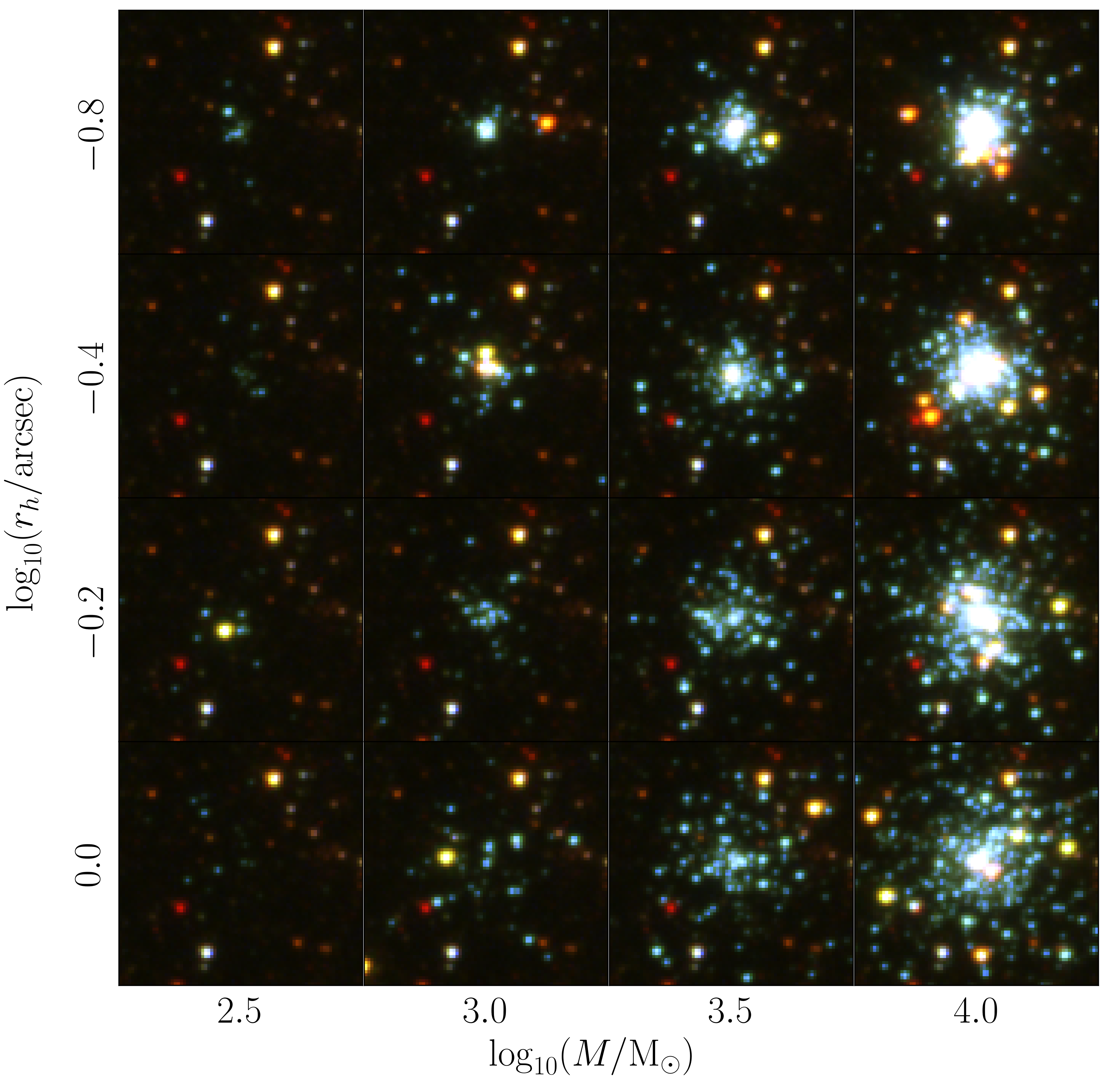} \\[0.5 pt]
        \small b)
    \end{tabular}
    \begin{tabular}{@{}c@{}}
        \includegraphics[width=0.48\textwidth]{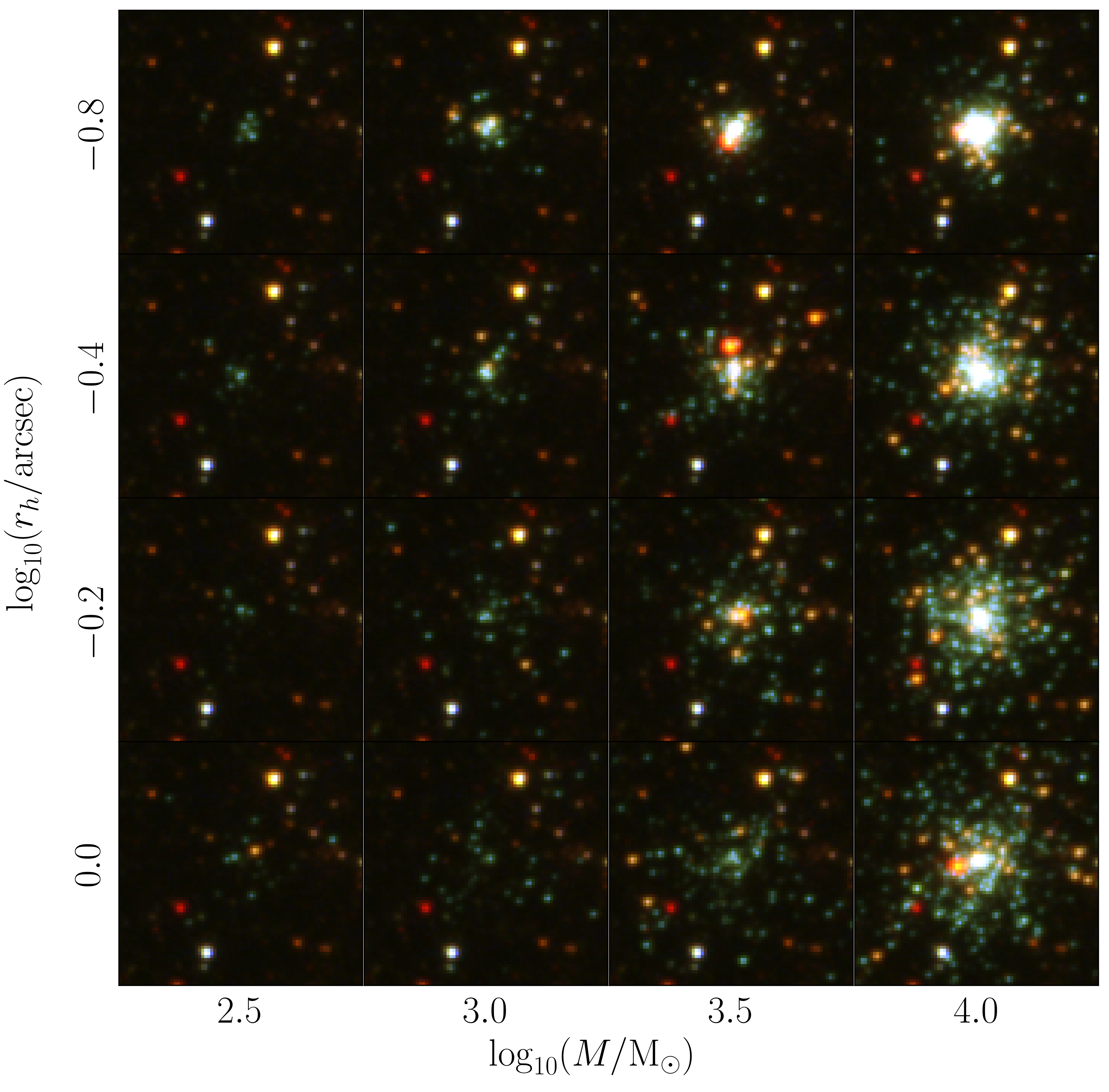} \\[0.5 pt]
        \small c)
    \end{tabular}
    \begin{tabular}{@{}c@{}}
        \includegraphics[width=0.48\textwidth]{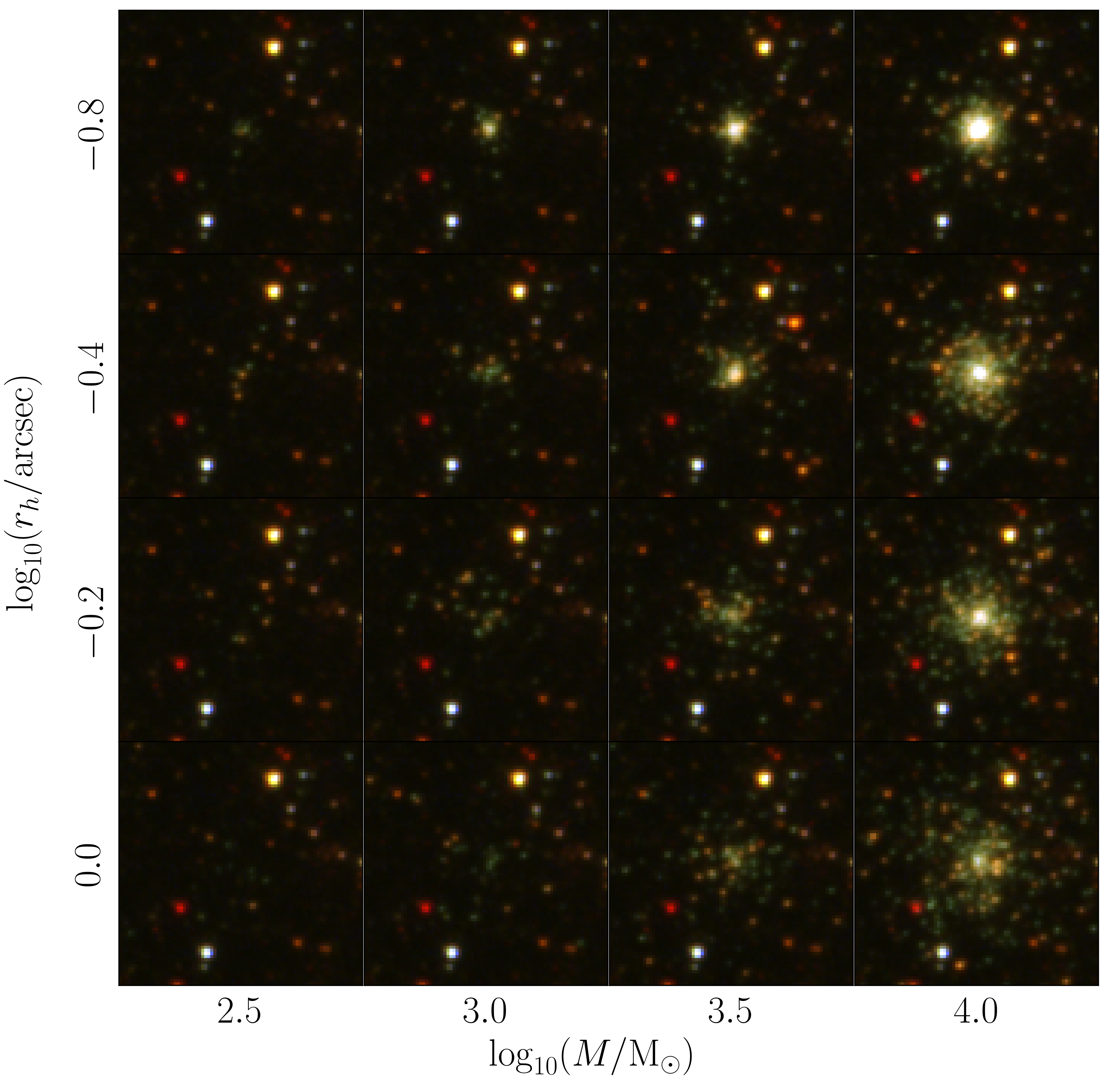} \\[0.5 pt]
        \small d)
    \end{tabular}
    \caption{Examples of generated clusters on a real background image. The ages of all of the displayed clusters are: a) $\log(t/{\rm yr})=7.0$, b) 8.0, c) 8.5, d) 9.0. The masses and $r_h$ values are varied as shown on the axes. The intensity scale of the images was normalized with the arcsinh function. The background is the same for all of the displayed clusters for clarity.}
    \label{fig:artificial_cluster_samples}
\end{figure*}

Examples of the generated clusters with different ages, masses, and sizes, covering most of the parameter space, are shown in Fig. \ref{fig:artificial_cluster_samples}, the background is the same for all of the displayed clusters.

We generated 200,000 images of artificial clusters as a training sample for our CNN. A batch size of 64 images per training step was used. To ensure that the training procedure of our CNN would converge, we experimented with various learning rates, starting from $\eta=0.1$, down to $\eta=0.0001$. The learning rate of $\eta=0.01$ gave the best performance on the validation set, so this was the value used for the final training of the network. A few different learning rate schedules were also tried out. Decreasing the learning rate twofold after every pass over the data gave the best results in our case. In order to control for overfitting we trained the network for 10 passes over the data while continuously monitoring its performance. The results on the validation set remained stable after about 3 passes, therefore, we chose to stop after 5 passes over the training data.

In addition, to test whether our training sample of 200,000 clusters is sufficient for the network to generalize, we also trained our network with as little as 50,000 and as many as 400,000 cluster samples. During these experiments we didn’t observe a significant difference between the inference results when trained with smaller or larger sample sizes.

\section{Results and discussion} \label{sec:results}
The data sets for testing the CNN were prepared by drawing from the same age and mass ranges as described in \S \ref{artificial_clusters}. However, for ease of analysis, the EFF model parameter $\gamma$ and $a$ pairs were chosen in such a way that the value of $r_h$ would be equal to one of the following values: 0.16, 0.25, 0.4, 0.63, 1.0 arcsec (see Fig. \ref{fig:eff_rad_construction}). The procedure of image creation was also identical to that of the training data. The backgrounds for clusters were picked making sure that they would not overlap with the backgrounds used for the training set.

\subsection{Parameter accuracy}
To test the performance of the CNN, we built a bank of 10,000 artificial clusters by varying all 3 cluster parameters. The ages were sampled from the range of $\log(t/{\rm yr})=[6.75, 9.25]$; mass was sampled from the range of  $\log(M/{\rm M_\odot})=[2.25, 3.75]$ and $r_h$, as shown in Fig. \ref{fig:eff_rad_construction}.

\begin{figure*}
    \centering
    \includegraphics[width=0.8\textwidth]{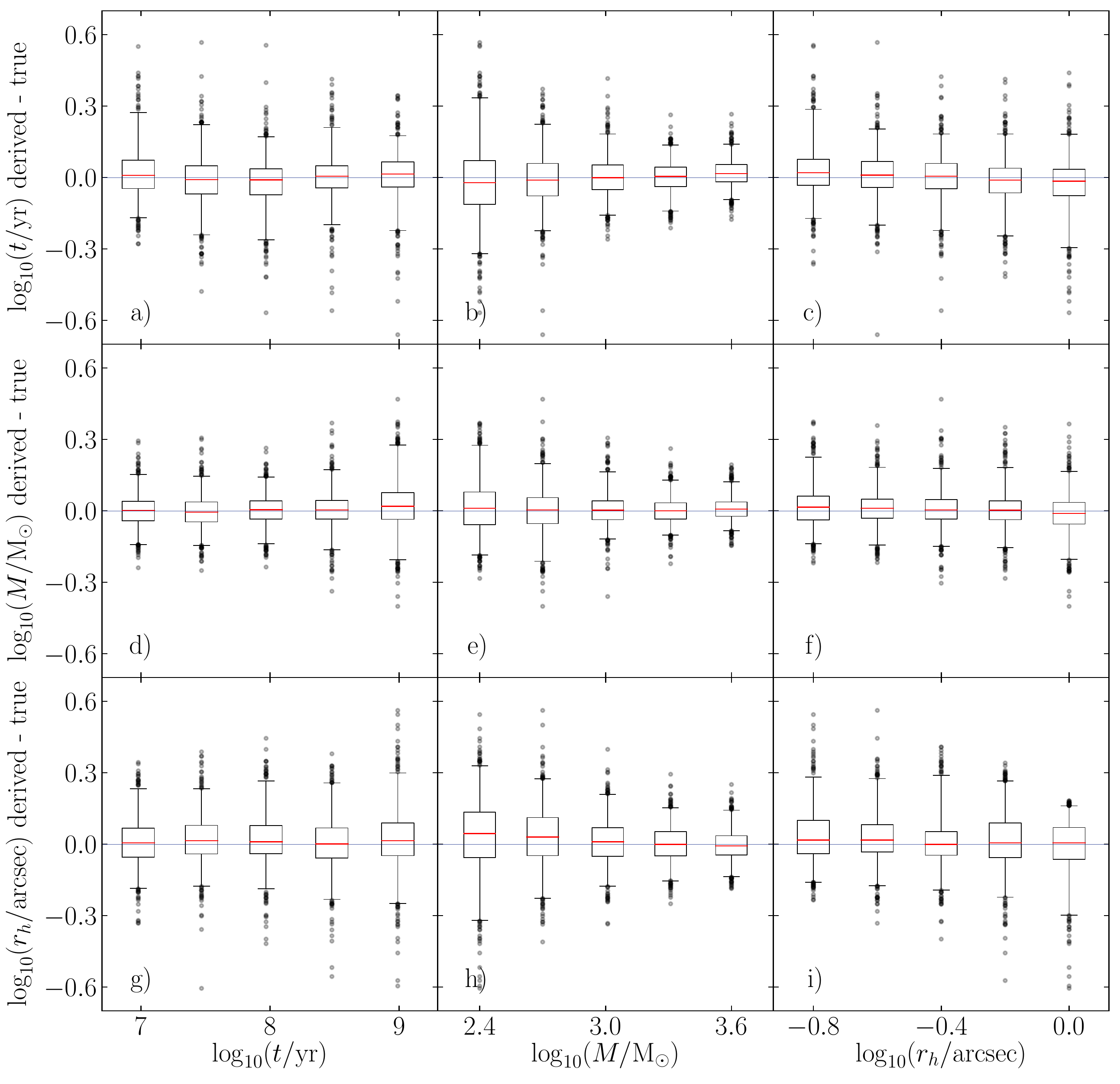}
    \caption{Differences between CNN derived and true parameters vs. true parameter (age, mass, and size) values of artificial clusters. The ages are binned into 0.5 dex width bins, and the masses into 0.3 dex width bins, while $r_h$ bins correspond to EFF model parameters indicated by dots on the constant $r_h$ values in Fig. \ref{fig:eff_rad_construction}. The widths of the boxes for age and mass correspond to half the widths of the bins. The spread of the parameter differences are displayed as box plots. The line in the middle of each box is the median error. Boxes extend from the 1st to the 3rd quartiles. The whiskers denote the range between the 2th and 98th percentiles. Anything above and below the whiskers is plotted as separate points.}
    \label{fig:artificial_comparison}
\end{figure*}

Differences between true and CNN-derived parameters vs. true parameter (age, mass and $r_h$) values of the testing set's artificial clusters are shown in Fig. \ref{fig:artificial_comparison}. For the comparison in the figure the true ages are binned into 0.5 dex width bins, and the true masses -- into 0.3 dex width bins. The $r_h$ value bins correspond to combinations of $a$ and $\gamma$ parameters denoted as dots in Fig. \ref{fig:eff_rad_construction}. The spread of the parameter differences within bins is displayed as box plots. The line in the middle of each box is the median difference between the true and derived values. Boxes extend from the 1st to the 3rd quartiles. Whiskers denote the range between the 2nd and the 98th percentiles. Anything above and below the whiskers is plotted as separate points.

Fig. \ref{fig:artificial_comparison}a shows no significant age difference between the true and derived values. The typical standard deviation of the age difference is estimated to be $\sim$0.1 dex. The youngest clusters show a slight systematic bias towards older ages, as can be seen by the non-symmetric whiskers. This is also true for the oldest clusters in the test sample; only the bias is reversed. This could be explained as an age boundary effect due to the stellar isochrone age range limits used for the CNN training. Fig. \ref{fig:artificial_comparison}b shows the true and derived age value differences plotted against the true cluster masses. The median differences have no systematic shifts. However, the least massive clusters show a larger mass error spread, with standard deviation of the differences as high as $\sim$0.15 dex. The errors get systematically lower as we move towards the higher masses, stabilizing at a standard deviation of $\sim$0.05 dex for the most massive clusters. This could be explained by the low signal-to-noise ratio of $\log_{10}(M/{\rm M_\odot})=2.4$ clusters, as can be seen from Fig. \ref{fig:artificial_cluster_samples}. No systematic errors are observed in Fig. \ref{fig:artificial_comparison}c, with the error standard deviations equal to $\sim$0.1 dex across the whole range of the true $r_h$ values.

Fig. \ref{fig:artificial_comparison}d shows no systematic errors when deriving cluster age as cluster mass varies. However, the spread of errors does get larger towards older clusters. In Fig. \ref{fig:artificial_comparison}e there are no systematic shifts; however errors are again larger for the lowest mass clusters. This could also be explained by the very low signal-to-noise ratio of $\log_{10}(M/{\rm M_\odot})=2.4$ clusters. In Fig. \ref{fig:artificial_comparison}f the derived mass values do not seem to vary with changing $r_h$ values.

Fig. \ref{fig:artificial_comparison}g shows consistently derived $r_h$ values with changing ages. Fig. \ref{fig:artificial_comparison}h shows systematically larger derived $r_h$ values for low mass clusters, as well as clearly larger errors. Fig. \ref{fig:artificial_comparison}i shows a clear systematic trend deriving larger $r_h$ values for small clusters, as well as lower values for large clusters. The shift in large clusters, with $r_h$ as high as 1.0 arcsec, could be explained by the fact that a significant portion of a cluster's stars do not fit into the 4$\times$4 arcsec images, the size of which is restricted by the chosen network architecture to minimize the background area for the majority of clusters.

\subsection{Star cluster stochastic effects}
In order to analyse the effects of star cluster stochasticity (initial mass function and star position sampling) on the CNN's inference results, we built a grid of clusters (200 per node) with fixed age, mass, and size parameters. The only random effects were background, star position, and mass sampling.

\begin{figure*}
    \centering
    \includegraphics[width=\textwidth]{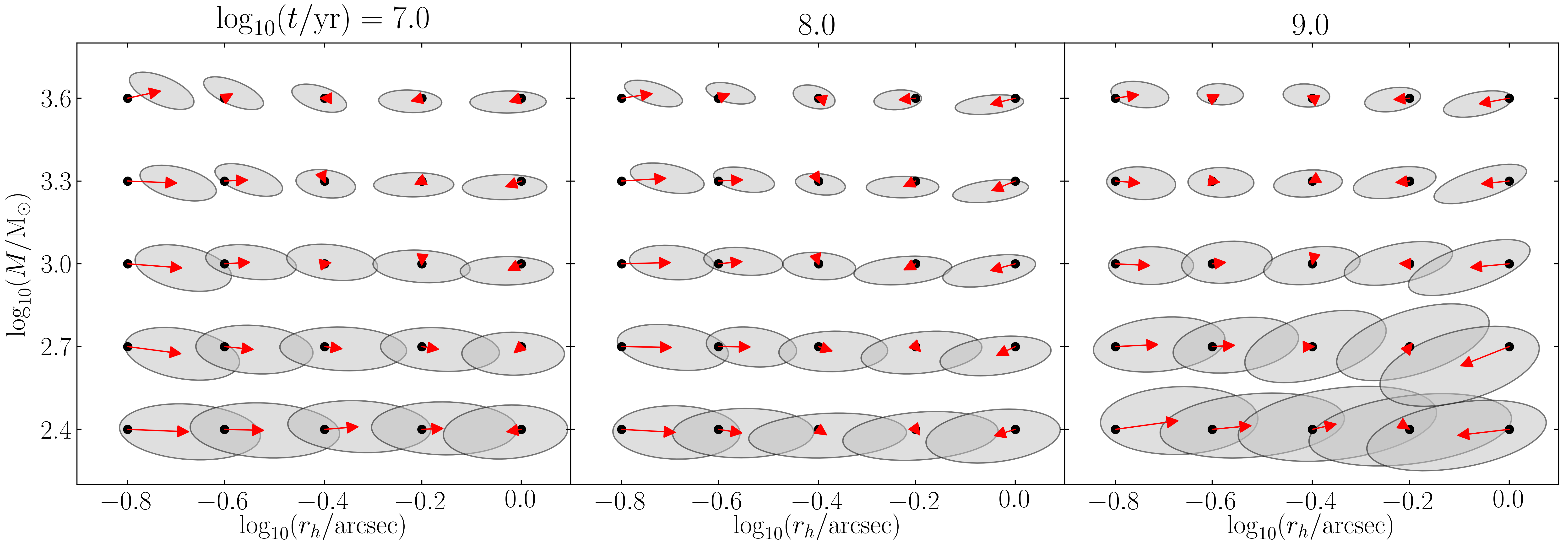}
    \caption{Tests results of CNN performance. Each black dot corresponds to the true parameters of 200 artificial clusters. The gray ellipses enclose one $\sigma$ of the inferred values (accuracy), with the red arrows pointing to the means of the distributions (biases). The panels show mass vs. $r_h$ for three different ages.}
    \label{fig:age_shifts}
\end{figure*}

Fig. \ref{fig:age_shifts} shows three panels with different ages: $\log_{10}(t/{\rm yr})=$ 7.0, 8.0 and 9.0. Each black dot corresponds to a node of 200 clusters of fixed true mass and $r_h$, as denoted on the axes. Ellipses show the spread of the derived parameter values of these nodes. The arrows show the shift of the mean of the values and the boundary of the ellipses enclose one $\sigma$ of the 2D distribution ($\sim$39\% of the clusters). The size of the ellipses at low masses are approximately three times larger than at high masses, especially so with the oldest ages. This is because of the low signal-to-noise ratio of the old low mass clusters, as can be seen in Fig. \ref{fig:artificial_cluster_samples}d. Low mass clusters tend to have slightly overestimated sizes. No significant correlated systematic shifts between the biases (red arrows) of derived mass and size can be seen.

\begin{figure*}
    \centering
    \includegraphics[width=\textwidth]{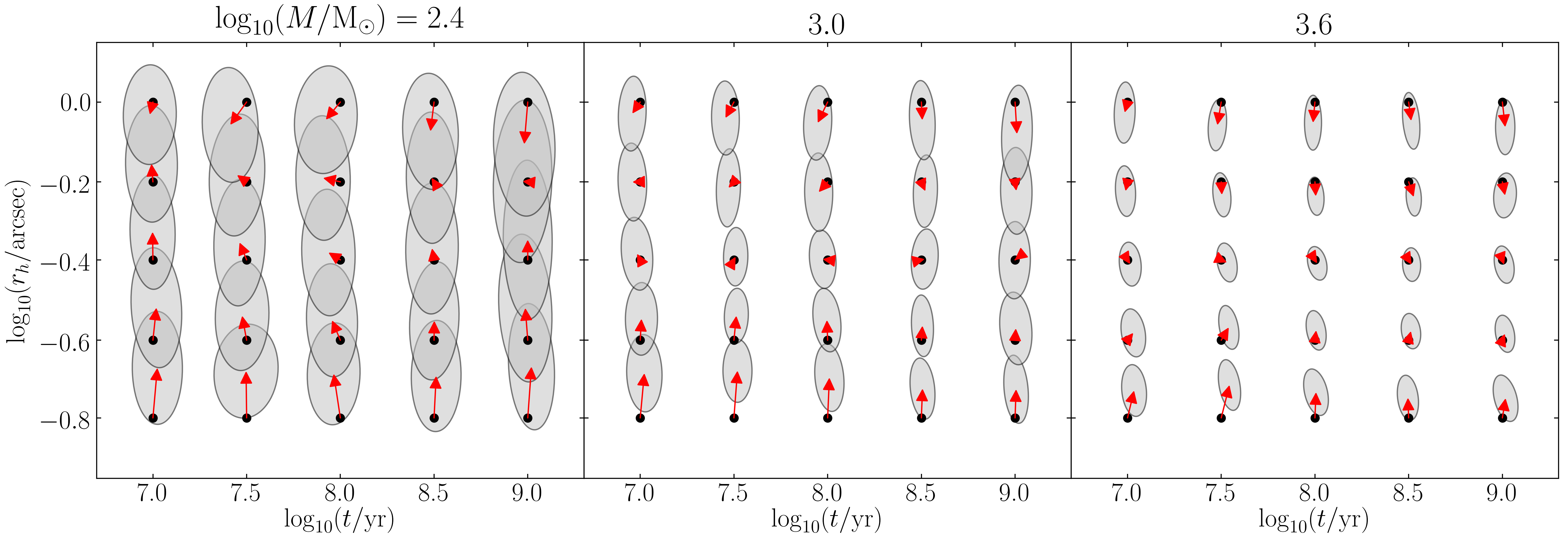}
    \caption{Same as Fig. \ref{fig:age_shifts}, but panels show $r_h$ vs. age for three different masses.}
    \label{fig:mass_shifts}
\end{figure*}

\begin{figure*}
    \centering
    \includegraphics[width=\textwidth]{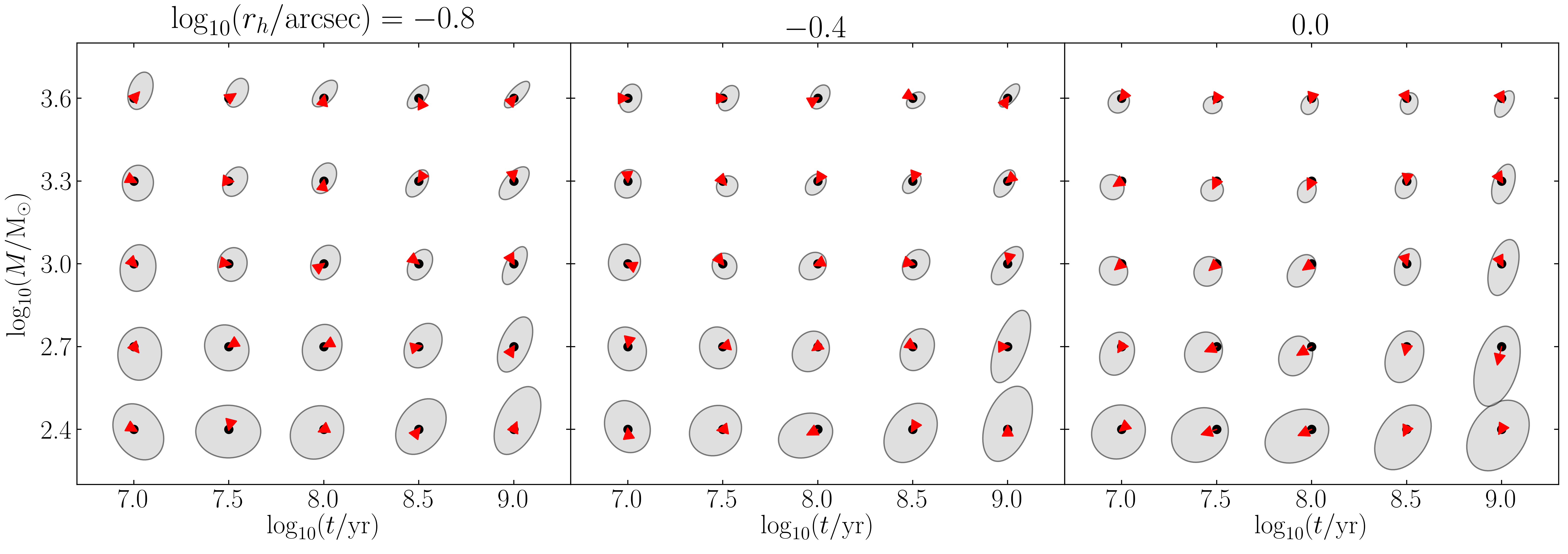}
    \caption{Same as Fig. \ref{fig:age_shifts}, but panels show mass vs. age for three different $r_h$ values.}
    \label{fig:rh_shifts}
\end{figure*}

Fig. \ref{fig:mass_shifts} shows three panels with different masses: $\log_{10}(M/{\rm M_\odot})=$ 2.4, 3.0, 3.6. Black dots mark nodes of age and size as denoted on the axes. The spread of the derived age and size values is largest for the lower mass clusters, being at least three times larger for $\log_{10}(M/{\rm M_\odot})=2.4$ than for $\log_{10}(M/{\rm M_\odot})=3.6$. However, at a fixed mass, there's no correlated systematic shift between the derived cluster's mass and size biases. The slight overestimation of sizes can be seen again for the lowest mass clusters.

Fig. \ref{fig:rh_shifts} shows three panels with different sizes: $\log_{10}(r_h/{\rm arcsec})=$ $-0.8$, $-0.4$, 0.0. Black dots mark nodes of age and mass as denoted on the axes. As seen in the previous figures, the spread of derived values tends to increase towards the lower mass clusters. However, there is no obvious influence of cluster size on the spread. A slight elongation of ellipses along the diagonal can be observed between the derived mass and age for the older grid clusters. This is because of the age-mass degeneracy, as the CNN has learned that older clusters have lower flux, and lower flux can be explained either by an older age or a lower mass. For $\log_{10}(t/{\rm yr})\lesssim8.0$ (as seen in Figs. \ref{fig:artificial_cluster_samples}a and \ref{fig:artificial_cluster_samples}b) stochastic effects are dominant; however for $\log_{10}(t/{\rm yr})>8.0$ (Figs. \ref{fig:artificial_cluster_samples}c and \ref{fig:artificial_cluster_samples}d) flux becomes the main factor in the mass-age correlation.

\subsection{Overall CNN performance} \label{sec:overall_performance}
The proposed CNN was verified on mock images of artificial clusters. It has demonstrated a high precision and no significant bias for semi-resolved clusters with ages between $\log_{10}(t/{\rm yr})=$ 7.0 and 9.0, and masses between $\log_{10}(M/{\rm M_\odot})=$ 2.4 and 3.6. Artificial cluster tests have demonstrated the effectiveness of CNNs in deriving star cluster parameters.

Even in the low mass regime it is possible to recover both age and mass as seen in Fig. \ref{fig:rh_shifts}. However, the plots show only one $\sigma$ ($\sim$39\%) of the 2D distributions, which means that for the lowest mass clusters five age groups can be identified. As mass increases, the number of identifiable groups increases. The same trend is also seen in Fig. \ref{fig:artificial_comparison}b and holds regardless of cluster size (see Figs. \ref{fig:artificial_comparison}c and \ref{fig:artificial_comparison}f). Note, that even though this is true on average, there are still outliers with a parameter derivation error as large as 0.6 dex for the low mass clusters (Fig. \ref{fig:artificial_comparison}b).

The low mass clusters shown in Fig. \ref{fig:artificial_cluster_samples}d are not visible to the naked eye. Nevertheless, the network still learns on the limited amount of signal and manages to produce valid parameter estimates. Even though the old age lowest mass clusters have been included, in real surveys they would be omitted because they would fall below the detection limit.

Meanwhile, for the cluster size, only three categories can be identified in the lowest mass regime as can be seen in Fig. \ref{fig:age_shifts}. This is, however, unsurprising as the signal-to-noise ratio of these clusters is low (see Fig. \ref{fig:artificial_cluster_samples}). This is especially true for the oldest clusters. However, higher masses allow us to identify up to five size groups. Fig. \ref{fig:artificial_comparison}h also confirms this finding.

Currently, evolutionary and structural parameters of semi-resolved clusters are estimated separately and in two completely different ways. Usually, for age and mass estimates a grid of models is constructed by varying the parameters of interest and comparing the resulting mock observations to real observations \citep{2008BaltA..17..337B, 2013A&A...550A..20D}. Meanwhile, a common approach for structural parameters is Markov chain Monte Carlo sampling of the parameter space \citep{2015BaltA..24..305N}. Both of these methods essentially sift through a large number of possible observations to get a likely parameter estimate or the distribution over the parameters.

CNNs, while still requiring a sample of mock clusters for their training, are able to learn the properties of the system they're modeling within their weights and generalize better to new examples. This also allows the derivation of both evolutionary and structural parameters in a unified way, using all of the available information in image pixels.

The information registered in image pixels is affected by many additional factors, which were not included in the CNN training; the PSF shape is expected to be the most significant among them for cluster size. In this paper, a fixed PSF, simulated for the center of the detector, was used. This was done for both the training cluster images and the test sets. In order to find the effects of variable PSFs on the network's performance, tests were done with the PSF taken from a corner of the detector; however the effects of this were negligible. In order to test the network's robustness to different views of the cluster, we also tried running the CNN on shifted and rotated cluster images. This, however, proved to have a negligible impact on the uncertainty of derived values.

Extinction and metallicity are also significant factors in age and mass derivation \citep{2014A&A...569A...4D, 2015A&A...574A..66D}. Since for the first time we attempt to derive both evolutionary and structural parameters, we assumed zero extinction and fixed metallicity just to demonstrate CNN performance. At fixed extinction and metallicity the CNN demonstrates good accuracy for age and mass derivation. This implies that the CNN can extract more information than integrated photometry due to its ability to learn the appearance of the cluster -- not only its integrated flux.

In comparison to methods where any sort of fitting or sampling procedure needs to be performed, a CNN can obtain parameter inference results from an image very efficiently. All experiments in this study were performed using a GeForce GTX 1080 graphics card. The initial training procedure on the 200,000 clusters takes $\sim$6 hours; however this only needs to be run once. Inference runs much faster, taking 40 seconds per 10,000 images of all three passbands.

\subsection{Tests on real clusters} \label{sec:real_clusters}
To validate our method on real star clusters we took the sample of clusters used by \cite{2017A&A...602A.112D}. From those clusters we selected only objects located on PHAT bricks 19-22. The CNN in our study was trained ignoring extinction entirely, therefore it is not possible to derive the parameters of significantly reddened clusters correctly. However, extinction estimates for the chosen clusters have been published by \cite{2017A&A...602A.112D}. In order to run our experiments, we corrected the cluster images in each passband for the extinction effect by increasing their corresponding flux.

\begin{figure*}
    \centering
    \includegraphics[width=0.8\textwidth]{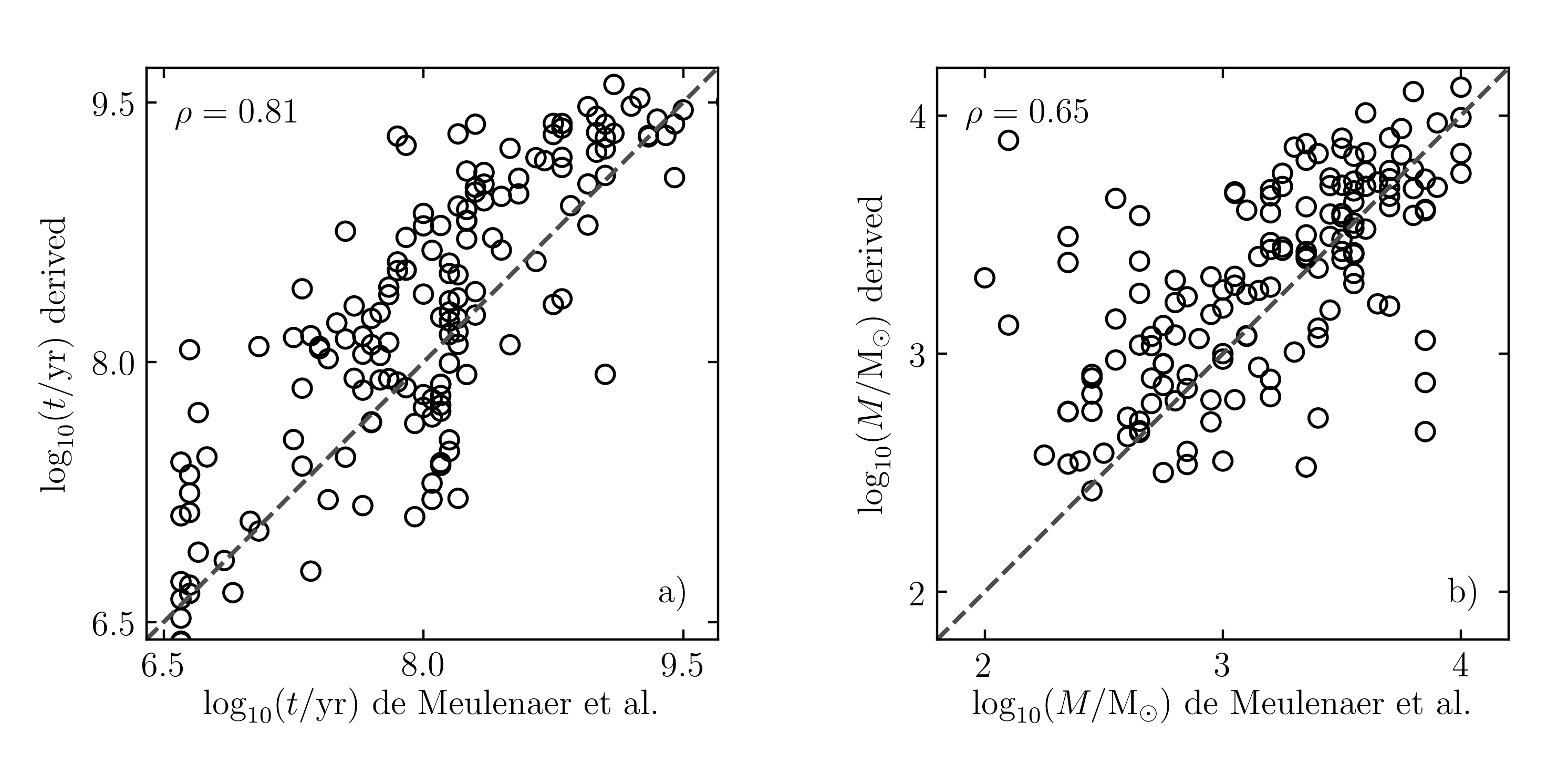}
    \caption{Comparison of ages (a) and masses (b) derived by \cite{2017A&A...602A.112D} and our CNN for 157 real PHAT clusters located in bricks 19-22. The correlation coefficient $\rho$ between values is displayed in the top-left of each subfigure.}
    \label{fig:real_clusters}
\end{figure*}

We then used our CNN to infer the cluster age and mass. Figure \ref{fig:real_clusters} shows a comparison between our results and those of \cite{2017A&A...602A.112D}. A good agreement between the derived values can be seen. However, this result can only be used for a preliminary validation of the method. By increasing the flux of our real cluster sample to remove the effects of extinction, we change the flux of not only the cluster itself, but also of its background. On the other hand, the artificial clusters used in this study, while themselves drawn without extinction, are superimposed on real backgrounds which are affected by extinction. This makes any results on real clusters unreliable, introducing possible degeneracies. However, the preliminary results are promising and show a clear applicability to real clusters.

In order to deal with real clusters correctly our CNN needs to be trained on images with various extinction levels, as well as to be able to predict extinction in the same way it currently predicts age, mass, and size. However, for a reliable derivation of extinction more photometrics filters are required. This work falls out of scope for this study, and is planned for the next paper in the series.

\section{Conclusions} \label{sec:conclusions}
We have proposed a convolutional neural network based on the ResNet architecture for simultaneous derivation of both evolutionary and structural parameters of star clusters from imaging data. Artificial cluster images were combined with real M31 backgrounds observed with the HST and used for training the neural network.

The proposed CNN was verified on mock images of artificial clusters. It has demonstrated a high accuracy and no significant bias for semi-resolved clusters with ages between $\log_{10}(t/{\rm yr})=$ 7.0 and 9.0, masses between $\log_{10}(M/{\rm M_\odot})=$ 2.4 and 3.6, and sizes between $\log_{10}(r_h/{\rm arcsec})=$ $-0.8$ and 0.0.

We have shown with artificial tests that CNNs can perform both structural and evolutionary star cluster parameter derivation directly from raw imaging data. This allows dealing with both unresolved and semi-resolved cases homogeneously, as well as utilizing multiple photometric passbands in an integrated manner.

\begin{acknowledgements}
This research was funded by a grant (No. LAT-09/2016) from the Research Council of Lithuania.

This research made use of Astropy, a community-developed core Python package for Astronomy (Astropy Collaboration, 2018).

Some of the data presented in this paper were obtained from the Mikulski Archive for Space Telescopes (MAST). STScI is operated by the Association of Universities for Research in Astronomy, Inc., under NASA contract NAS5-26555.

We are thankful to the anonymous referee who helped improve the paper.
\end{acknowledgements}
\bibliography{library}
\bibliographystyle{aa}
\end{document}